\documentclass[onecolumn]{ametsocV6.1}

\providecommand{\appendix}{\par\setcounter{section}{0}\setcounter{subsection}{0}\renewcommand{\thesection}{Appendix~\Alph{section}}}

\usepackage{graphicx}

\usepackage{tikz-cd}
\usepackage{comment}
\usepackage{subcaption}

\usepackage{pgfplots}
\usepgfplotslibrary{colormaps}

\pgfplotsset{compat=newest} 

\newcommand\EN           {\end{equation}}
\newcommand\bes           {\begin{subequations}}
\newcommand\esu           {\end{subequations}}

\newcommand{\bu}{\boldsymbol{\mathrm{u}}}

\newcommand{\csf}{\mathsf{c}}

\newcommand{\U}{\boldsymbol{\mathrm{U}}}
\newcommand{\A}{\boldsymbol{\mathrm{A}}}
\newcommand{\x}{\boldsymbol{\mathrm{X}}}
\newcommand{\Q}{\boldsymbol{\mathrm{Q}}}
\newcommand{\Ell}{\mathscr{L}}

\title{Instability of the halocline at the North Pole}

\authors{Christian Puntini \correspondingauthor{Christian Puntini, christian.puntini@univie.ac.at}}

\affiliation{Faculty of Mathematics, University of Vienna, \\Oskar-Morgenstern-Platz 1, 1090, Vienna, Austria.}

\abstract{In this paper, we address the issue of stability for the near-inertial Pollard waves, as a model for the halocline in the region of the Arctic Ocean centred around the North Pole, derived in \cite{mio}. Adopting the short-wavelength instability approach, the stability of such flows reduces to studying the stability of a system of ODEs along fluid trajectories, leading to the result that, when the steepness of the near-inertial Pollard waves exceeds a specific threshold, those waves are linearly unstable. The explicit dispersion relation of the model allows us to easily compute such a threshold, knowing the physical properties of the water column.}

\begin{document}

\maketitle

\section{Introduction}
Exact solutions, albeit being quite rare, are fundamental tools in the study of (geophysical) fluid flows. Even if they typically describe ideal conditions that may not correspond with the actual complexities of the observed fluid dynamics, they can provide the basis for more directly relevant analyses: small perturbations can be superimposed on the existing exact solution to describe more complex flows. The primary obstacle to obtaining explicit solutions in geophysical fluid dynamics lies in the nonlinear nature of the governing equations, which renders direct analytical computation highly challenging. The first explicit exact solution of these nonlinear equations was discovered by Gerstner in 1809 for a homogeneous fluid \citep{gerstner}. This solution was later rediscovered by \cite{Froude}, \cite{Rankine}, and \cite{Reech}, and subsequently extended to heterogeneous fluids by \cite{Dubreil}.

A key feature of Gerstner’s solution—and of the subsequent extensions—is that it is formulated in the Lagrangian framework. Lagrangian descriptions offer the advantage of explicitly characterising the fluid kinematics (see \citep{Bennett}), since they track the motion of individual fluid particles (in contrast with the Eulerian framework, in which the fluid flow is described by specifying the velocity field at fixed spatial locations).

The Lagrangian formalism has proved particularly effective for deriving exact solutions of several oceanographic models. \cite{Pollard} generalized Gerstner’s trochoidal solution to incorporate Earth’s rotation, and more recently, Gerstner-like constructions have been employed to model equatorial waves \citep{C2012, C-GRL, C2013, C2014, 25}, wave–current interactions \citep{17, 43, 44}, and edge waves \citep{Mollo, 7, 32, Weber, Miao, S2014}. See also \citep{S2013} for Gerstner waves in the case of a still-water surface, and \citep{31} for Pollard-type waves in equatorial regions.\\
Once an exact solution is given, a natural question that arises is its stability. Broadly speaking, a flow is stable when small perturbations do not alter the main characteristics of the motion, whereas instability occurs when the effects of some disturbances grow with time (see \cite{CG, stability}). A detailed understanding of the instability is key to identifying the mechanisms that prompt the transition from the exact coherent large-scale structure to turbulent, chaotic flow, which in turn causes mixing (see \cite{Staquet_CISM, S_S}).\\
The aim of this paper is to present a stability analysis of a three-dimensional geophysical flow derived in \citep{mio} to describe the halocline of the region of the Arctic Ocean centred around the North Pole via a $3-$layer model. We adopt the short-wavelength instability approach (developed by \cite{E_S_78}, \cite{Bayly}, \cite{FriedlanderVishik1991} and \cite{LifschitzHameiri}), which is particularly suited for flows described in the Lagrangian framework. In the seminal work of \citet{LeblancJFM} it has been applied to Gerstner's rotational free-surface gravity waves, and than it has been used to prove instability of many models in oceanography (see \citet{CG, IK2015, IK2016, Kruse, Kruse2,  Chu_IK_Y,  ChuIKY, Miao}). \\
The paper is structured as follows: Section \ref{flow} provides an overview of the fundamental characteristics of the Arctic halocline and the mathematical model employed, before establishing the governing equations that describe the flow dynamics of interest, and reviewing the nonlinear solution developed in \citep{mio} that describes the Arctic halocline via near-inertial Pollard waves (for the central region of the Arctic Ocean, around the North Pole).
Section \ref{instability} contains the main result of this work: the instability of the near-inertial Pollard waves describing the halocline. Using the short-wavelength instability approach (adapted from \citet{Kruse}, where it has been used in the context of Pollard waves for the first time), stability is studied by analysing a system of ODEs along the fluid particle paths, and a criterion for instability is established: if the wave steepness exceeds a certain threshold, the waves become unstable. The explicit dispersion relation of the solution allows for straightforward computation of the numerical values of this threshold, given the physical properties of the water column.
Finally, we conclude with a discussion of the results in Section \ref{discussion}.

\section{Description of the flow and governing equations}\label{flow}
\begin{figure}
    \centering
    \includegraphics[width=.7\linewidth]{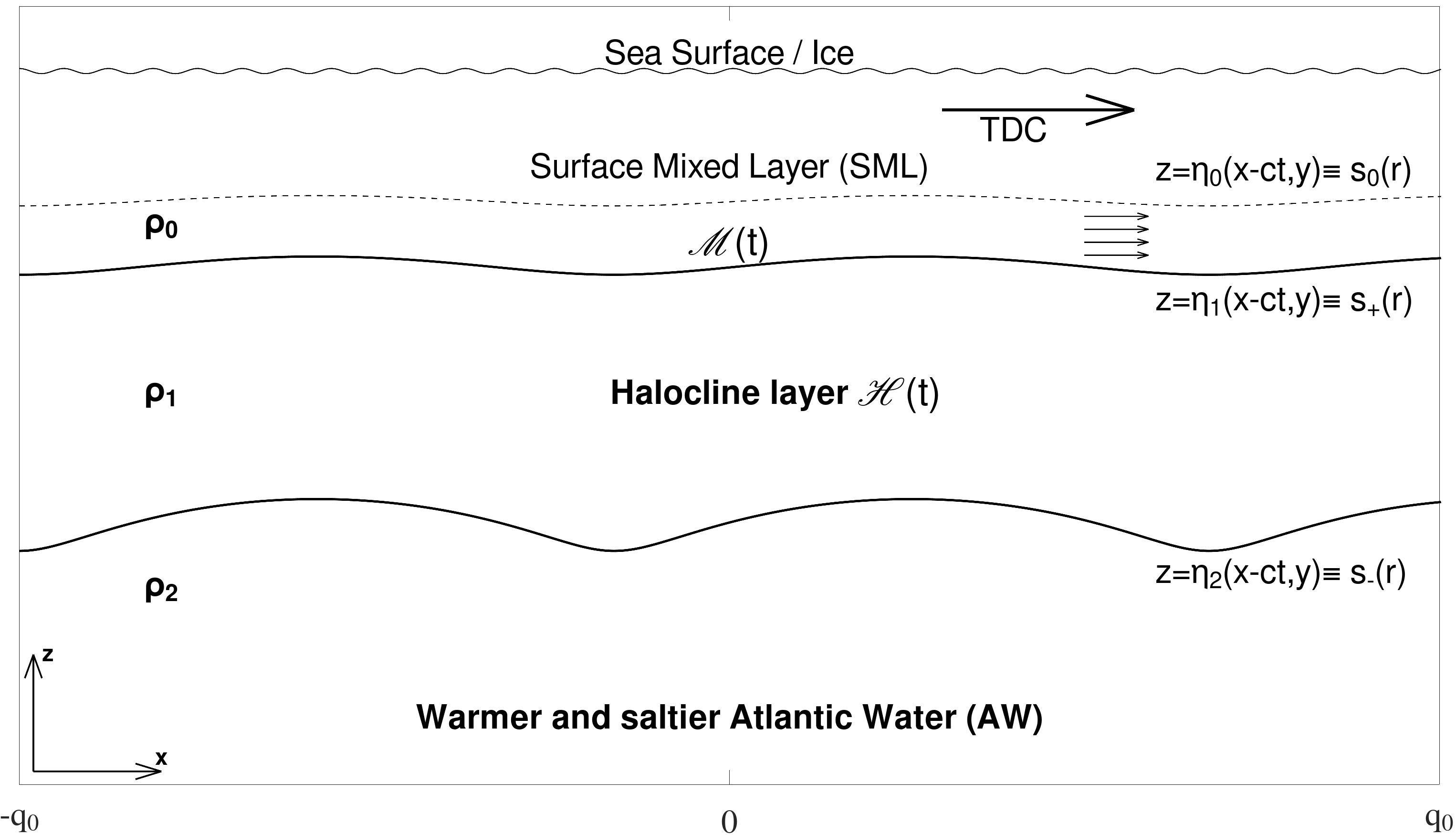}
    \caption{A depiction of the model we are considering developed in \cite{mio}, consisting of 3 layer: the surface mixed layers (of density $\rho_0$) at which bottom the mean surface current is present, the halocline layer (of density $\rho_1$) and the bottom layer consisting of Atlantic Water (AW) with density $\rho_2$. The halocline layer is described by nonlinear near-inertial Pollard waves with amplitude increasing with depth, which induce a wave motion at the bottom of the surface layer. This picture is for illustrative purposes and is not to scale.}
    \label{strati}
\end{figure}

\begin{figure}[ht]
\centering
\includegraphics[width=0.5\textwidth]{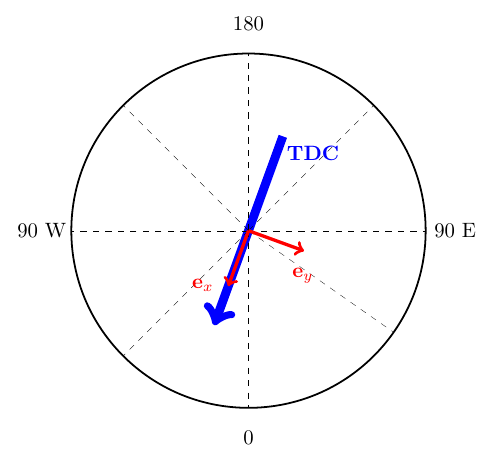}
\caption{Depiction of the basis at the North Pole, with $\mathbf{e}_x$ aligned with the Transpolar Drift Current and $\mathbf{e}_y$ perpendicular to it. The $\mathbf{e}_z$-axis, as usual, points upward.}
\label{Basis NorthPole}
\end{figure}

The Arctic halocline represents a region of pronounced stratification that prevents---in the central region of the Arctic Ocean---sea ice from thermally interacting with the warmer, more saline Atlantic Water located in the deeper portions of the water column. The Arctic Ocean exhibits a distinctive vertical structure comprising several distinct layers. The surface mixed layer consists of cold, fresh water extending from 5 to 100 meters in depth. Below this lies the halocline, which extends to depths of 40-200 meters, underlain by a layer of relatively warm, saline Atlantic Water \citep{48, 42}. These layer boundaries vary seasonally with temperature and sea ice: winter cooling and expanded ice deepen the halocline, while summer warming and ice retreat shift it toward the surface \citep{48}. The deepest stratum consists of Arctic Deep Water. As this deepest layer is not relevant to our model, we focus exclusively on the uppermost three layers. Moreover, it is worth remarking that variations in the structure of the Arctic Ocean, especially in its central part, have been established as indicators of the global warming (see e.g. \cite{Morison2018, 51}).\\
Physical processes within the mixed layer, primarily ice motion driven by wind forcing and the Transpolar Drift Current (TDC), generate significant flow at the base of the mixed layer. The TDC represents the dominant Arctic Ocean current that transports surface waters and sea ice from the Laptev Sea and East Siberian Sea toward Fram Strait at velocities of approximately $0.07 \mathrm{ms^{-1}}$ in the vicinity of the North Pole. These processes collectively induce currents with magnitudes of approximately $0.1 \mathrm{ms^{-1}}$  at the lower boundary of the mixed layer (see \cite{Guthrie}).\\
Although the halocline exhibits regional variations—including the cold halocline layer in the Eurasian Basin, the Pacific Halocline Waters in the Amerasian Basin, and the lower halocline water \citep{42}—the simplified model employed herein assumes constant densities within each layer under consideration, as we are considering a relatively small region of the Arctic Ocean centred around the North Pole. This assumption is also justified by the minimal density variations observed within individual strata \citep{51}. Our model incorporates three constant densities: $\rho_0$, $\rho_1$ and  $\rho_2$, where  $\rho_0<\rho_1<\rho_2$. Here,  $\rho_0$ represents the fresh, cold water of the surface mixed layer above the halocline,  $\rho_1$ denotes the density of halocline water, and  $\rho_2$ corresponds to the density of the saltier, warmer Atlantic Water (AW) below the halocline (see  Fig. \ref{strati}).\\
With increasing depth, the influence of surface turbulence diminishes progressively, particularly in the presence of ice cover (which is a characteristic of the central part of the Arctic Ocean, subject of our investigation). Consequently, we postulate the existence of a potentially thin layer immediately above the lower boundary of the Surface Mixed Layer (SML), denoted as $\eta_1$, wherein the fluid exhibits unidirectional flow aligned with the Transpolar Drift Current (TDC) at a mean velocity of approximately $0.1\,\mathrm{ms^{-1}}$. The upper boundary of this layer is designated as $\eta_0$ (the specific functional form of $\eta_0$ does not influence our theoretical framework).
Furthermore, we adopt the assumption that the layer beneath the halocline remains essentially motionless, with its lower boundary denoted as $\eta_2$, representing the bottom interface of the halocline. See Fig. \ref{strati}.\\
Having established our model framework for the region of the Arctic Ocean around the North Pole, we now proceed to briefly set forth the governing equations. Away from the boundary layers, friction and viscous effects are negligible, allowing the fluid to be treated as ideal \citep{Maslowe, 13}. Consequently, the momentum equations governing the fluid motion take the form of the Euler equations, and, due to the small variations of the Coriolis parameter in high-latitude regions (note that $\sin90^\circ=1$ and that $\sin80^\circ\approx0.985$), together with the Earth’s oblate, pole-flattened geometry,
we can adopt an $f$-plane approximation for the study of flows in regions centred at the North Pole and that extend southward to $80^\circ$ N. In our setting, the $\mathbf{e}_x$ axis is aligned with the Transpolar Drift Current and the axis $\mathbf{e}_y$ is perpendicular to it (see Fig. \ref{Basis NorthPole}). The $\mathbf{e}_z$-axis, as usual, points upward. It is worth noting that such $f$-plane approximation is not built starting from the governing equations in the classical spherical coordinates, but in the ``rotated" spherical coordinate system developed in \citep{14} suitable for the North Pole (where the classical spherical coordinates fail). We refer to \citep{mio} for a more in-detail exposition of the governing equations (see also \citep{mioGoverning} for an in-depth derivation of the governing equation in this new rotated coordinate system).\\
Therefore, the Euler equations in the $f$-plane approximation read as
\begin{equation}\label{Euler f plane artic}
\left\{\begin{aligned}
	&U_t + U U_x + V U_y + W U_z- fV=
	-\frac{1}{\rho}
	P_x, \\
	&V_t+ U V_x+ V V_y +W V_z +
	f U=
	-\frac{1}{\rho}
	P_y, \\
	&W_t+U W_x+V W_y + WW_z+g
	=
	-\frac{1}{\rho}
	P_z 	,
\end{aligned}\right.
\end{equation}
where  $f=2\Omega\approx1.46\cdot 10^{-4}\, \mathrm{s^{-1}}$ and $(U,V,W)=:\U$ are the velocity component associated to the basis $\left(\mathbf{e}_x, \mathbf{e}_y, \mathbf{e}_z\right)$, and pedices $x,\, y,\, z,\, t$ indicate partial derivatives. The density $\rho$ in \eqref{Euler f plane artic} will be set equal to $\rho_0,\ \rho_1$ or $\rho_2$ depending on the layer of the fluid we are considering.\\
Moreover, the seawater is treated as incompressible, as a change of about $500\, \mathrm{kPa}$ in pressure leads to only a $0.025\%$ change in the water density (see \citep{Maslowe}), giving that the incompressibility condition
\begin{equation}\label{cont eq}
U_x+V_y+W_z=0,
\end{equation}
must hold, as well as the mass conservation
\begin{equation}\label{Cmass}
    \rho_t+U\rho_x+V\rho_y+W\rho_z=0.
\end{equation}
As we are considering constant densities, \eqref{Cmass} is always satisfied.\\
In addition to the governing equations \eqref{Euler f plane artic} and \eqref{cont eq} we require the solution to satisfy the dynamic boundary condition (i.e. the pressure must be continuous across the surfaces $\eta_1$ and $\eta_2$), and the kinematic boundary condition
\begin{equation}\label{KBC}
W=\eta_{i,t}+U\eta_{i,x}+V\eta_{i,y} \qquad \text{on}\ z=\eta_i,\quad i=1,2,
\end{equation}
preventing the mixing of fluid particles between different layers (see \citep{ConstantinBook}).\\
As we are assuming the bottom part of the portion of Arctic Ocean under consideration to be in hydrostatic state, we have that $u=v=w=0$ for $z<\eta_2(x-\csf t,y)$. Consequently, the continuity equation \eqref{cont eq} is satisfied, while the equations of motion \eqref{Euler f plane artic} reduce to
\begin{equation}
	P_x=	
	P_y= 0,\quad P_z	=-\rho_2 g.
\end{equation}
Those can be easily integrated, leading to the following expression for the pressure (where $P_2$ is a constant)
\begin{equation}\label{pressure under halocline}
P(x,y,z)=P_2-\rho_2 g z.
\end{equation}
\\
In the halocline layer (namely, $\eta_2(x-\csf t,y)\leq z \leq \eta_1(x-\csf t,y)$) situated above the motionless layer just described, and which denote by $\mathscr{H}(t)$, we look for an explicit solution in the Lagrangian formalism of \eqref{Euler f plane artic}, \eqref{cont eq} and \eqref{KBC}, fulfilling also the dynamic boundary conditions (that is, the pressure must be continuous across $\eta_1$ and $\eta_2$). At each time $t$ we specify the position $\x=(x,y,z)$ of the fluid particles in terms of the labeling/material variables $(q, r, s)$ by writing
\begin{equation}\label{Lag}
\left\{	\begin{aligned}
	&x=q-\frac{ma}{k}e^{-ms}\sin\Theta,\\
	&y=r+\frac{fma}{k^2\csf}e^{-ms}\cos\Theta,\\
	&z=-d_0+s-ae^{-ms}\cos\Theta,
\end{aligned}\right.
\end{equation}
where $k=\frac{2\pi}{L}>0$ is the wave number corresponding to the wavelength $L$, and we set $a>0$. Moreover, we denoted 
\begin{align}
    \Theta=k(q-\csf t),\\
     m=\sqrt{\frac{k^4\csf^2}{k^2\csf^2-f^2}}.\label{m}
\end{align}
The material variables are chosen so that
\begin{equation}
(q,r,s) \in [- \mathtt{q}_0, \mathtt{q}_0]\times[- \mathtt{r}_0, \mathtt{r}_0]\times [\mathrm{s}_-(r),\mathrm{s}_+(r) ],
\end{equation}
where $s=\mathrm{s}_-(r)\geq \mathrm{s}^*>0$ represents $\eta_2$, $s=\mathrm{s}_+(r)>\mathrm{s}_-(r)$ represents $\eta_1$, and $d_0-\mathrm{s_-}$ is the mean depth of the halocline base. Moreover, $\mathtt{q}_0 $ and $ \mathtt{r}_0$ are not in excess of $1000$ km, so that for the area under consideration the hypothesis of our model hold (namely a unidirectional mean flow (the TDC) is present, and the water columns is approximated by a 3-layer ocean) away from the lands and from other flows (such as the Beaufort gyre). Since the determinant of the Jacobian of the map \eqref{Lag}, which reads as
\begin{equation}\label{18}
\begin{aligned}
\left(\frac{\partial(x,y,z)}{\partial(q,r,s)}\right):=\begin{pmatrix}
	x_q & y_q & z_q \\
	x_r & y_r & z_r	\\	
	x_s & y_s & z_s	
\end{pmatrix}=\begin{pmatrix}
	1-ma\,e^{-ms}\cos\Theta & -f\frac{ma}{k\csf}e^{-ms}\sin\Theta & kae^{-ms}\sin\Theta \\
	0 & 1 & 	0\\
	\frac{m^2a}{k}e^{-ms}\sin\Theta & -f\frac{m^2a}{k^2\csf}e^{-ms}\cos\Theta &1+mae^{-ms}\cos\Theta
\end{pmatrix},
\end{aligned}
\end{equation}
is non-zero and time-independent, being equal to
\begin{equation}\label{det}
\Delta=1-m^2a^2\,e^{-2ms}>0
\end{equation}
the flow is incompressible (see \citep{ConstantinBook, 44}) and \eqref{cont eq} is satisfied. The velocity and acceleration of a particle can be computed using \eqref{Lag}, giving, respectively
\begin{equation}\label{20}
\left\{\begin{aligned}
	&U=\frac{	Dx}{Dt}=ma\csf e^{-ms}\cos\Theta,\\
	&V=\frac{	Dy}{Dt}=\frac{fma}{k}\, e^{-ms}\sin\Theta,\\
	&W=\frac{	Dz}{Dt}=-k\csf a\, e^{-ms}\sin\Theta,
\end{aligned}\right.
\end{equation}
and
\begin{equation}\label{21}
\left\{\begin{aligned}
	&\frac{	D U}{Dt}=k\csf^2 ma\, e^{-ms}\sin\Theta,\\
	&\frac{	D V}{Dt}=-fma\csf \, e^{-ms}\cos\Theta,\\
	&\frac{	D W}{Dt}=k^2\csf^2 a\, e^{-ms}\cos\Theta,
\end{aligned}\right.
\end{equation}
which lead to the following expression for the pressure in the halocline layer $\mathscr{H}(t)$, obtained by integrating \eqref{Euler f plane artic} coupled with \eqref{20} and \eqref{21} :
\begin{equation}\label{25}
\begin{aligned}
	P(q-\csf t, r,s )=P_1+\rho_1\left\{ \frac{k^2\csf^2 a^2}{2}e^{-2ms}+a \left[m\left(\csf^2-\frac{f^2}{k^2}\right)+g\right]e^{-ms}\cos\Theta-gs\right\}.
\end{aligned}
\end{equation}
Lastly, we denote by $\mathscr{M}(t)$ the layer above the halocline, bounded below by the halocline upper surface $\eta_1$ and above by $\eta_0$. The oscillations of the halocline induces the wave-motion of this stratum, and - as physical measurements highlights the presence of a mean current of about $0.1\, \mathrm{ms^{-1}}$ in the direction of the Transpolar Drift Current \citep{Guthrie} - we include such feature by assuming a uniform horizontal current in the $x$-direction (in our coordinate system aligned with the TDC), denoted by $\csf_0$. Adopting again the Lagrangian formalism, the position $\x=(x,y,z)$ of the fluid particles at time $t$ in terms of the material variables $(q,\ r,\ s)$ is given by
\begin{equation}\label{LagA}
\left\{	\begin{aligned}
	&x=q-\frac{ma}{k}e^{-ms}\sin\Theta-\csf_0 t,\\
	&y=r+\frac{fma}{k^2\csf}e^{-ms}\cos\Theta,\\
	&z=-d_0+s-ae^{-ms}\cos\Theta.
\end{aligned}\right.
\end{equation}
The labeling variables in this case are chosen so that
\begin{equation}
(q,r,s) \in [- \mathtt{q}_0, \mathtt{q}_0]\times[- \mathtt{r}_0, \mathtt{r}_0]\times [\mathrm{s}_+(r),\mathrm{s}_0(r) ],
\end{equation}
where $s=\mathrm{s}_0(r)$ represents the upper surface $\eta_0$ of the layer $\mathscr{M}(t)$, and, as before $s=\mathrm{s}_+(r)$ represents $\eta_1$, with $\mathtt{q}_0, $ and $\mathtt{r}_0$ chosen as previously.\\
The Jacobian of the map \eqref{LagA} is the same as the one of \eqref{Lag}, hence the flow described by \eqref{LagA} satisfies the incompressibility condition \eqref{cont eq}.
The velocity  of a particle in this layer is given by
\begin{equation}\label{20 e 21A}
\left\{\begin{aligned}
	&U=\frac{	Dx}{Dt}=ma\csf \, e^{-ms}\cos\Theta-\csf_0,\\
	&V=\frac{	Dy}{Dt}=\frac{fma}{k}\, e^{-ms}\sin\Theta,\\
	&W=\frac{	Dz}{Dt}=-k\csf a\, e^{-ms}\sin\Theta,
\end{aligned}\right.
\end{equation}
whereas the acceleration is equal to \eqref{21}. Computing the mean Lagrangian velocity in the $x-$direction (that we recall, is aligned with the TDC) over a wave period $T=\frac{L}{\csf}$, we get
\begin{equation}\label{meanLagM}
         \langle U\rangle_L =\frac{1}{T}\int_0^T U(q-\csf t,r,s)=-\csf_0,
    \end{equation}
    therefore, as one assumption of our model is to have a mean flow in the direction of the TDC, we need to set $-\csf_0>0$, that is $\csf_0<0$. See Fig. \ref{strati} for a depiction of the solution.\\
    As for the halocline layer, we integrate  the governing equations \eqref{Euler f plane artic} (using \eqref{20 e 21A} and \eqref{21}) to obtain the following expression for the pressure
\begin{equation}\label{25A}
\begin{aligned}
	P(q-\csf t, r,s )=P_0+\rho_0 \left\{ f \csf_0 r +\frac{k^2\csf^2a^2}{2}e^{-2ms}+ a \left[g+m\csf^2 +m\frac{f^2}{k^2}\left(\frac{\csf_0}{\csf}-1\right)\right] e^{-ms}\cos\Theta-gs\right\}.
\end{aligned}
\end{equation}
For every $r\in[- \mathtt{r}_0, \mathtt{r}_0]$, the upper surface of the halocline $\eta_1$ is described by $s=\mathrm{s}_+(r)$, while the halocline base $\eta_2$ is determined by setting $s=\mathrm{s}_-(r)$. The continuity of the pressure is imposed by equating \eqref{25} and \eqref{25A} at $\eta_1$ (i.e. for $s=\mathrm{s}_+(r)$) and by equating \eqref{pressure under halocline} and \eqref{25} at $\eta_2$ (i.e. for $s=\mathrm{s}_-(r)$), giving
\begin{equation}\label{a}
\begin{aligned}
\text{at}\ \eta_1:& \quad \left\{\begin{aligned}
	&        P_0-P_1=\left(\frac{\rho_1-\rho_0}{2}\right) k^2\csf^2a^2e^{-2m\mathrm{s}_+}-(\rho_1-\rho_0)g\mathrm{s}_+ -\rho_0 f \csf_0 r ,\\
	&   \rho_1\left[m\left(\csf^2-\frac{f^2}{k^2}\right)+g\right]= \rho_0\left[m\left(\csf^2-\frac{f^2}{k^2}\right)+g+m\frac{f^2}{k^2}\frac{\csf_0}{\csf}\right],
\end{aligned}\right.\\ 
\text{at}\ \eta_2:& \quad \left\{\begin{aligned}
	&       P_2-P_1=\frac{1}{2}\rho_1 bk\csf(bk\csf+df)e^{-2m\mathrm{s}_-}+(\rho_2-\rho_1)g\mathrm{s}_- -\rho_2 g d_0,\\
	&      m\left(\csf^2-\frac{f^2}{k^2}\right)=\frac{\rho_2-\rho_1}{\rho_1} g.
\end{aligned}\right.    
\end{aligned}
\end{equation}
The existence and uniqueness of $\mathrm{s}_+(r)$ and $\mathrm{s}_-(r)$ (i.e. $\eta_1$ and $\eta_2$) can be established using the implicit function theorem on the first and third equations in \eqref{a} (for the details, see \citep{mio}). The second equation in \eqref{a}, using  the expression for $m$ in \eqref{m},  can be rewritten as
\begin{equation}\label{condition}
    (\rho_1-\rho_0)\left(g+|\csf|\sqrt{k^2\csf^2-f^2}\right)=\rho_0\frac{f^2|\csf|}{\sqrt{k^2\csf^2-f^2}}\frac{\csf_0}{\csf},
\end{equation} providing the following two conditions that have to be satisfied
\begin{equation}
    \csf^2>\frac{f^2}{k^2},\qquad\text{and}\qquad
    \rho_0\frac{f^2|\csf|}{\sqrt{k^2\csf^2-f^2}}>0.
\end{equation}
Defining the reduced gravity $\mathfrak{g}$ as
\begin{equation}\label{redg}
\mathfrak{g}:=\left(\frac{\rho_1-\rho_0}{\rho_0}\right) \frac{\rho_2}{\rho_1}g,
\end{equation}
and coupling the second and forth equations in \eqref{a} lead to the dispersion relation
\begin{equation}\label{dispersionC}
\csf=-\sqrt{\frac{f^2}{k^2}+\frac{f^4\csf_0^2}{\mathfrak{g}^2k^2}}.    
\end{equation}
    Since $\mathfrak{g}$ is of order of $10^{-2}\,\mathrm{ms^{-2}}$ to $ 10^{-4}\,\mathrm{ms^{-2}}$ (see the discussion in Section \ref{discussion}), $f=2\Omega\approx 1.46\cdot10^{-4}\,\mathrm{s^{-1}}$ and $\csf_0\lesssim 0.1\,\mathrm{ms^{-1}}$, it follows that 
    \begin{equation}
        \frac{\frac{f^2}{k^2}}{\frac{f^4\csf_0^2}{\mathfrak{g}^2k^2}}=\frac{\mathfrak{g}^2}{f^2\csf_0^2}>>1,
    \end{equation}
    hence
\begin{equation}
\csf^2\approx \frac{f^2}{k^2},
\end{equation}
namely the modulus of the period of the wave $T=\frac{L}{\csf}$ is approximately $\frac{2\pi}{f}=T_{\rm i}$, where $T_{\rm i}$ is the inertial period of the Earth, so the wave motion described in \eqref{Lag} and \eqref{LagA} is near-inertial.  As pointed out by \citet{19,20}, near-inertial waves are the most energetic ones in the ocean.\\
Finally, we recall one essential interesting feature of the model under consideration. It has been observed that the upper surface of the halocline becomes shallower in the Eurasian Basin and deeper in the Amerasian Basin (see \cite{51}). This feature is captured by the presented solution. In fact, evaluating the first equation in \eqref{a} at $s=\mathrm{s}_+(r)$, and then differentiating with respect to $r$ gives
\begin{equation}
    \mathrm{s}'_+(r)=\frac{\rho_0}{(\rho_0-\rho_1)}\frac{f\csf_0}{(mk^2\csf^2a^2e^{-2m\mathrm{s}_+(r)}+g)},
\end{equation}
so we obtain that $\mathrm{s}'_+(r)>0$ with $\csf_0<0$ (which is an assumption of our model), which is exactly the expected behaviour when moving from the Amerasian to the Eurasian Basin.

\section{Instability analysis}\label{instability}
\begin{figure}
    \centering
    \includegraphics[width=.7\linewidth]{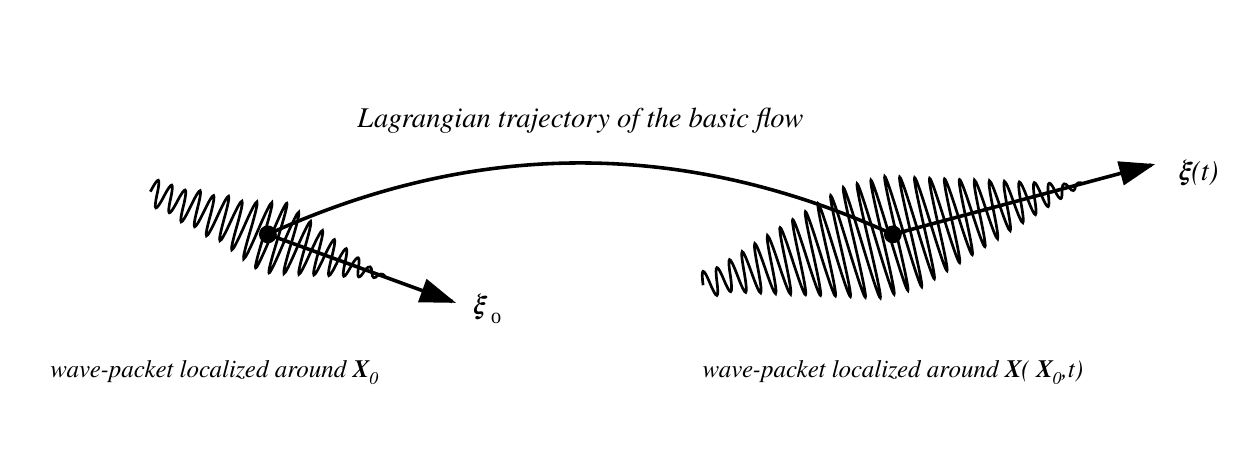}
    \caption{Evolution of a high-frequency disturbance along the basic flow. At $t=0$ a small wave packet of the form \eqref{initial condition} is localized near $\x_0$, with amplitude $\A_0$ and $\boldsymbol{\xi}(0)=\boldsymbol{\xi}_0$. It is advected by the basic flow $\U$ along the Lagrangian trajectory given by the first equation in \eqref{leading order}, while its amplitude $\A(\x,t)$ and the vector $\boldsymbol{\xi}(t)$ evolve due to local shear and strain according to the last two equations in \eqref{leading order}. At time $t$ the packet is localized near $\x(\x_0,t)$ with amplitude $\A(\x,t)$ and vector $\boldsymbol{\xi}(t)$; during this process, $\A$ may grow or decay depending on the properties of the basic flow $\U$ .}
    \label{instabilityIMAGE}
\end{figure}

This section is devoted to the main result of the present work: a criterion for the onset of instability for the model proposed in \cite{mio} and briefly reported in the previous section, based on the wave steepness of the solution in \eqref{Lag} and \eqref{LagA}. To this end, we first recall the short-wavelength instability approach for Pollard waves developed in a different setting by \cite{Kruse} (see also \cite{Kruse2}), tailoring it to our model. We then apply it to the near-inertial Pollard waves for the 3-layer model describing the arctic halocline under consideration. Differently to other works, the simple dispersion relation \eqref{dispersionC} will allow to express the instability criterion relatively to the ocean properties.\\
Instability is associated with the way (small) perturbations evolve and possibly grow in time when superimposed on a given basic flow. In particular, the onset of instability is characterized by the amplification of infinitesimal disturbances, whose dynamics determine whether the basic state is stable or unstable.
Small perturbations $\bu(\x,t)$, $p(\x,t)$ of the basic flow $\U(\x,t),\ P(\x,t)$ are governed by the following set of equations, obtained by linearizing the equations of motion \eqref{Euler f plane artic} and \eqref{cont eq}:
\begin{equation}\label{linearized}
\begin{aligned}
    \bu_t+(\U\cdot\nabla)\bu+(\bu\cdot \nabla)\U +\Ell \bu=-\frac{1}{\rho}\nabla p,\qquad
    \nabla\cdot\bu=0,
\end{aligned}
\end{equation}
where
\begin{equation}
    \Ell=\begin{pmatrix}
        0&-f&0\\
        f&0&0\\
        0&0&0
    \end{pmatrix},
\end{equation}
and, for the analysis of the flow, we will set $\rho=\rho_0$ or $\rho=\rho_1$, depending on the layer under consideration.\\
The short-wavelength instability approach consists in the study of the evolution in time of the solutions of the linearized system \eqref{linearized} in the following WKB form
\begin{equation}\label{up}
    \begin{aligned}
        \bu(\x,t)\approx\left[\A(\x,t)+\epsilon\hat{\A}(\x,t)\right]e^{\frac{i}{\epsilon}\Phi(\x,t)},\qquad
        p(\x,t)\approx\epsilon\mathcal{P}(\x,t)e^{\frac{i}{\epsilon}\Phi(\x,t)},
    \end{aligned}
\end{equation}
where $\epsilon$ is a small parameter, $\A$ and $\hat{\A}$ are vector functions, while $\Phi$ and $\mathcal{P}$ are scalar functions, and with the initial condition
\begin{equation}\label{initial condition}
    \bu_0:=\bu(\x,0)=\A(\x,0)e^{\frac{i}{\epsilon}\Phi(\x,0)}
\end{equation}
representing a sharply-peaked initial disturbance. Substituting the expression for $\bu$ and $p$ given by \eqref{up} into the linearized equations \eqref{linearized}, we obtain the following system governing the evolution of the short-wavelength perturbation at leading order 
   \begin{align}
   & \Phi_t+\U\cdot\nabla\Phi=0, \label{251}\\
   &     \A_t+(\U\cdot\nabla)\A+(\A\cdot\nabla)\U+\Ell A=\frac{\nabla\Phi\cdot[2(\A\cdot\nabla)\U+\Ell \A]}{||\nabla\Phi||^2}\nabla\Phi\label{252},
    \end{align}
    consisting of an eikonal equation and a transport equation, and with the initial conditions
\begin{equation}
    \Phi(\x,0)=\Phi_0(\x),\qquad \A(\x,0)=\A_0(\x)
\end{equation}
satisfying 
\begin{equation}
      \A_0\cdot\nabla\Phi_0=0.
\end{equation}
In equation \eqref{252} $||\cdot||$ denotes the standard $L^2$-norm. Setting $\boldsymbol{\xi}=\nabla\Phi$, and taking the gradient of \eqref{251} we have that
\begin{equation}
    \boldsymbol{\xi}_t+(\U\cdot\nabla)\boldsymbol{\xi}+(\nabla\U)^T\boldsymbol{\xi}=0,
\end{equation}
namely
\begin{equation}\label{xi}
\frac{D\boldsymbol{\xi}}{Dt}=-(\nabla\U)^T\boldsymbol{\xi}
\end{equation}
with the initial condition $\boldsymbol{\xi}(0)=\boldsymbol{\xi}_0$ and where $\nabla\U$ is the velocity gradient tensor,  given by
\begin{equation}\label{nablaU2}
 \nabla \U=\begin{pmatrix}
	U_x & U_y & U_z\\
	V_x & V_y & 	V_z\\
	W_x & W_y & W_z
\end{pmatrix}.
\end{equation}
Moreover, writing $\boldsymbol{A}=(A_1,A_2,A_3)$, equation \eqref{252} reads as
\begin{equation}
  \frac{D\A}{Dt}=-(\A\cdot\nabla)\U-\Ell \A+\frac{\boldsymbol{\xi}\cdot[2(\A\cdot\nabla)\U+\Ell \A]}{||\boldsymbol{\xi}||^2}\boldsymbol{\xi}
  \end{equation}
  with the initial condition $\A(0)=\A_0$.\\
Summing up, as the trajectory of $\U$ through a point $\x_0$ is given by (see \citet{Bennett})
\begin{equation}
     \frac{D\x}{Dt}=\U(\x,t),\qquad \text{with}\qquad        \x(0)=\x_0,
    \
\end{equation}
we obtain that the evolution in time of $\x$, $\boldsymbol{\xi}$ and $\A$, at leading order, is given by the following coupled system of ODEs
  \begin{equation}\label{leading order}
    \left\{\begin{aligned}
      &  \frac{D\x}{Dt}=\U(\x,t),\\
        &\frac{D\boldsymbol{\xi}}{Dt}=-(\nabla\U)^T\boldsymbol{\xi},\\
        &   \frac{D\A}{Dt}=-(\A\cdot\nabla)\U-\Ell \A+\frac{\boldsymbol{\xi}\cdot[2(\A\cdot\nabla)\U+\Ell \A]}{||\boldsymbol{\xi}||^2}\boldsymbol{\xi},
    \end{aligned}\right.
\end{equation}
with associated initial conditions 
\begin{equation}\label{ic}
    \x(0)=\x_0, \qquad \boldsymbol{\xi}(0)=\boldsymbol{\xi}_0, \qquad \A(0)=\A_0,
\end{equation}
satisfying the constraint
\begin{equation}\label{constraint}
    \A_0\cdot \boldsymbol{\xi}_0=0.
\end{equation}
The first equation in \eqref{leading order} describes the particle trajectory of the basic flow, while the second and third govern - at leading order - the evolution of the local wave vector and the amplitude of the perturbation along the particle trajectory, respectively (see Fig. \ref{instabilityIMAGE}).\\
The growth rate of $\A$ is analogous to the concept of Lyapunov exponent (see e.g. \cite{Friedlander, CG, Henry}). If fore some initial position $\x_0$ we have that 
\begin{equation}
    \Lambda(\x_0)=\limsup_{t\rightarrow\infty}\frac{1}{t}\ln\left(\sup_{\substack{|\boldsymbol{\xi}_0|=|\A_0|=1\\
   \boldsymbol{\xi}_0\cdot \A_0=0} }|\A(\x,t ; \boldsymbol{\xi}_0,\A_0)|\right)>0
\end{equation}
then, for a given fluid trajectory, the particles separate at an exponential rate, and the fluid flow is unstable \citep{Friedlander}.\\
We now apply the set of equations \eqref{leading order} (with initial conditions \eqref{ic} and \eqref{constraint}) to the nonlinear wave motion described in Section \ref{flow} to prove its instability. The basic flows $\U$ considered are therefore  those in \eqref{Lag} and \eqref{LagA} (in the halocline and in the layer above it, respectively). In this case, the velocity gradient tensor (which is the same in the two layers $\mathscr{H}(t)$ and $\mathscr{M}(t)$), is given by
\begin{equation}\label{nablaU}
\begin{aligned}
 \nabla \U&=\begin{pmatrix}
	U_x & U_y & U_z\\
	V_x & V_y & 	V_z\\
	W_x & W_y & W_z
\end{pmatrix} = \begin{pmatrix}
	U_q & U_r & U_s\\
	V_q & V_r & 	V_s\\
	W_q & W_r & W_s
\end{pmatrix}  \begin{pmatrix}
	q_x & r_x & s_x \\
	q_y & r_y & s_y\\
	q_z& r_z & s_z		
\end{pmatrix}=\\
&=\frac{1}{\Delta}\begin{pmatrix}
-ma\csf k \,e^{-ms}\sin\Theta	 &0 & m^2a\csf\,e^{-ms}(ma\,e^{-ms}-\cos\Theta) \\
	 fma\, e^{-ms}(\cos\Theta+ma\, e^{-ms})&0 &-\frac{fm^2a}{k}e^{-ms}\sin\Theta \\
	-k^2\csf a\, e^{-ms}(\cos\Theta+ma\,e^{-ms})& 0 & 		km\csf a \, e^{-ms}\sin\Theta
\end{pmatrix},
\end{aligned}
\end{equation}
where $\Delta=1-m^2a^2e^{-2ms}$ is the determinant in \eqref{det} and we made use of the following expression for the inverse of \eqref{18}
\begin{equation}\label{18inverse}
\begin{aligned}
 \left(\frac{\partial(q,r,s)}{\partial(x,y,z)}\right)&=\left(\frac{\partial(x,y,z)}{\partial(q,r,s)}\right)^{-1}=\begin{pmatrix}
	q_x & r_x & s_x \\
	q_y & r_y & s_y\\
	q_z& r_z & s_z		
\end{pmatrix}=\\
&=\frac{1}{\Delta}
\begin{pmatrix}
	{1+ma\,e^{-ms}\cos\Theta}& \frac{fma\,e^{-ms}\sin\Theta}{k\csf} & -kae^{-ms}\sin\Theta\\
	0 & \Delta & 	0\\
	-\frac{m^2a\,e^{-ms}\sin\Theta}{k} & \frac{fm^2a\,e^{-ms}\cos\Theta-m^3a^2f\, e^{-2ms}}{k^2\csf} &{1-mae^{-ms}\cos\Theta}
\end{pmatrix},  
\end{aligned}
\end{equation}
as well as the following expressions for the $q$, $r$ and $s$ derivatives of \eqref{Lag} and \eqref{LagA}:
\begin{equation}\label{velocityDerivatives}
\begin{aligned}
	&U_q=-kam\csf\, e^{-ms}\sin\Theta,\\
	&V_q=fma\, e^{-ms}\cos\Theta,\\
	&W_q=-k^2\csf a\, e^{-ms}\cos\Theta,
\end{aligned} \qquad
\begin{aligned}
	&U_r=0,\\
	&V_r=0,\\
&W_r=0,
\end{aligned}
\qquad\begin{aligned}	&U_s=-am^2\csf\, e^{-ms}\cos\Theta,\\
	&V_s=-f\frac{am^2}{k}\, e^{-ms}\sin\Theta,\\
	&W_s=km\csf a\, e^{-ms}\sin\Theta.
\end{aligned}
\end{equation}
From \eqref{nablaU}, it is immediate to see that
\begin{equation}\label{relation}
 U_x=-W_z,\qquad   V_x=-\frac{fm}{k^2\csf}W_x\qquad\text{and}\qquad  V_z=-\frac{fm}{k^2\csf}W_z.
\end{equation}

\begin{figure}
    \centering
    \begin{subfigure}{0.49\textwidth}
       \centering
    \includegraphics[width=\linewidth]{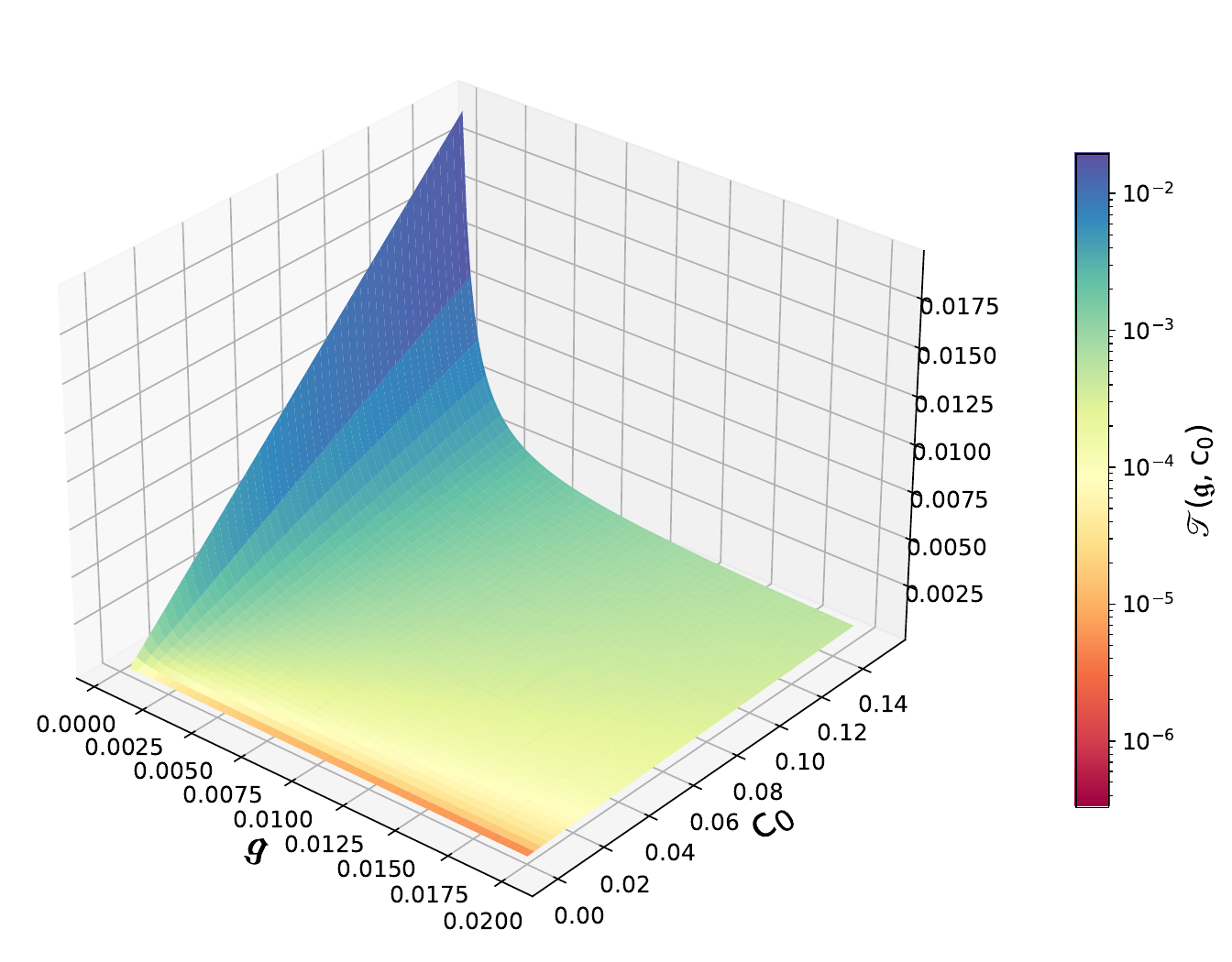}
    \caption{$\mathfrak{g}\in [5\cdot10^{-4},2\cdot10^{-2}] $.}
    \label{fig:threshold}
    \end{subfigure}
    \hfill
    \begin{subfigure}{0.49\textwidth}
      \centering
    \includegraphics[width=\linewidth]{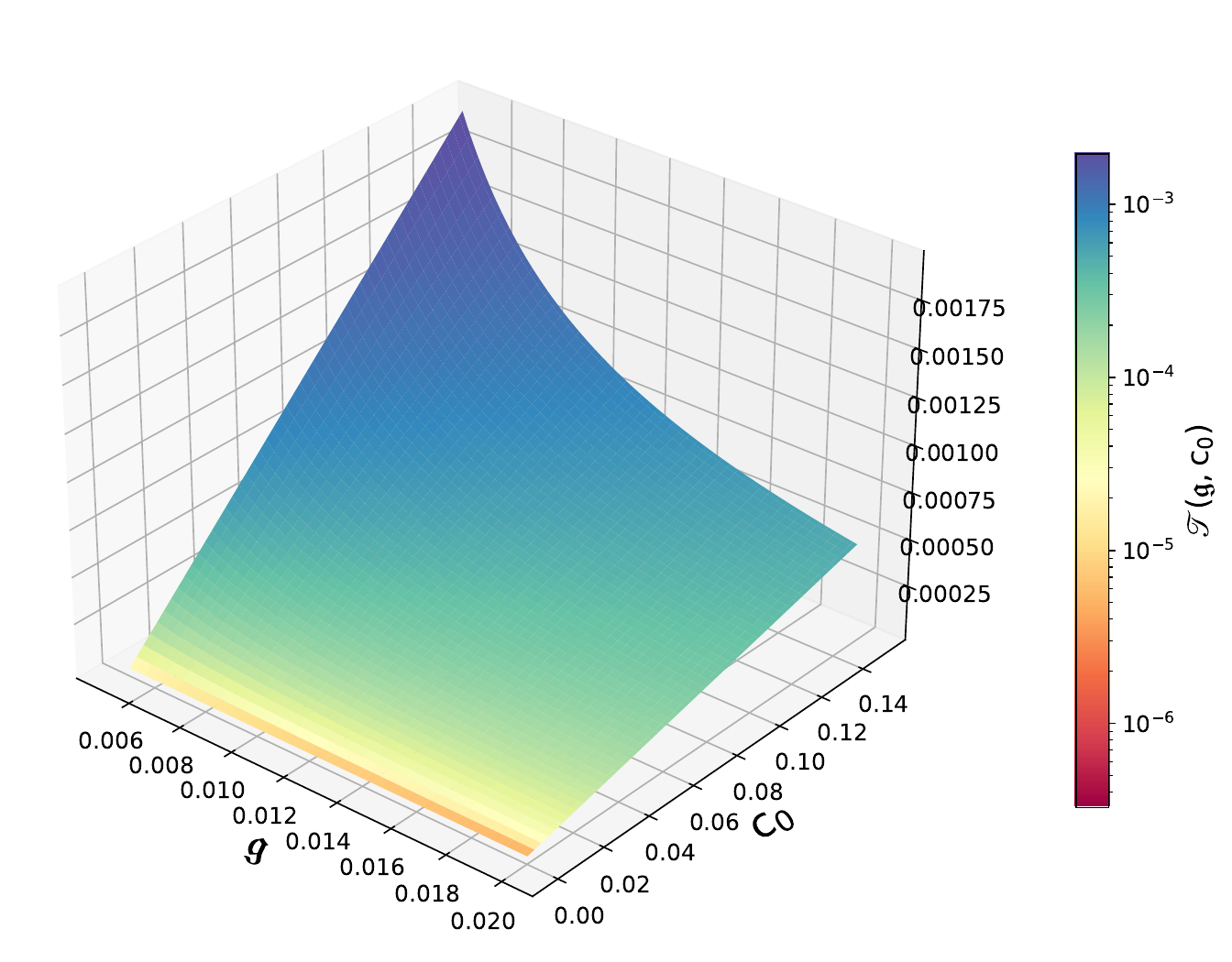}
    \caption{$\mathfrak{g}\in [5\cdot10^{-3},2\cdot10^{-2}] $.}
    \label{fig:threshold2}
    \end{subfigure}    
    \caption{Graph of the threshold $\mathscr{T}$ as a function of $\mathfrak{g}$ and $\csf_0$, with $\mathfrak{g}\in [5\cdot10^{-4},2\cdot10^{-2}] $ (left) and $\mathfrak{g}$ limited to $\mathfrak{g}\in [5\cdot10^{-3},2\cdot10^{-2}] $ (right). In both pictures, $\csf_0\in[0, 0.15]$.}
    \label{fig:both_figures}
\end{figure}

In order to prove the instability of the Pollard waves in \eqref{Lag} and \eqref{LagA} it is not necessary to investigate the system \eqref{leading order} for all the initial data. Instead, it is sufficient to provide one initial disturbance resulting in an exponential growth of $\A$. With this scope, we choose the following initial wave vector 
\begin{equation}
    \boldsymbol{\xi}_0=\left(0,\, \frac{{k}}{m},\, \frac{f}{k\csf}\right),
\end{equation}
and it can be proven that a solution to $\frac{D\boldsymbol{\xi}}{Dt}=-(\nabla\U)^T\boldsymbol{\xi}$ is given by  (see \citep{Kruse} for the details)
\begin{equation}\label{XI}
    \boldsymbol{\xi}(t)=\begin{pmatrix}
        \Xi_{11} & \Xi_{12} & \Xi_{13}\\
        0 & 1 & 0\\
          \Xi_{31} & \Xi_{32} & \Xi_{33}
    \end{pmatrix}  \boldsymbol{\xi}_0
\end{equation}
where
\begin{equation}
  \begin{aligned}
        \Xi_{11}&=1+ma\, e^{-ms}(\cos\Theta-\cos(kq))-m^2a^2\, e^{-2ms}\cos(k\csf t),\\
        \Xi_{12}&=\frac{f}{k\csf}\left[ma\, e^{-ms}(\sin\Theta-\sin(kq))-m^2a^2\,e^{-2ms}\sin(k\csf t)\right],\\
        \Xi_{13}&=ka\, e^{-ms}(\sin(kq)-\sin\Theta)+mka^2\, e^{-2ms}\sin(k\csf t),\\
        \Xi_{31}&=\frac{m^2 a}{k}e^{-ms}(\sin(kq)-\sin\Theta)-\frac{m^3a^2}{k}e^{-2ms}\sin(k\csf t),\\
        \Xi_{32}&=\frac{f}{k^2\csf}\left[m^3a^2e^{-2ms}(\cos(k\csf t)-1)+m^2a e^{-ms}(\cos\Theta-\cos(kq)\right],\\
        \Xi_{33}&=1-m^2a^2e^{-2ms}\cos(k\csf t)-ma\, e^{-ms}(\cos\Theta-\cos(kq)),
    \end{aligned}
\end{equation}
from which it follows that 
\begin{equation}\label{xiT}
    \boldsymbol{\xi}(t)=\left(0,\, \frac{{k}}{m},\, \frac{f}{k\csf}\right)\qquad \forall\, t\geq 0.
\end{equation}
Inserting the expression \eqref{nablaU} for $\nabla\U$ and the above one for $\boldsymbol{\xi}(t)$ into the third equation of \eqref{leading order}, which governs the time evolution of the amplitude of the disturbance $\A$, we get
\begin{equation}\label{50}
\begin{aligned}    
  \frac{D\A}{Dt}
 & =
 \underbrace{\begin{pmatrix}
     -U_x& f & -U_z\\
     \frac{fm}{k^2\csf}W_x-\frac{f^3}{k^2\csf^2} & 0 &  \frac{fm}{k^2\csf}W_z\\
      -W_x +\frac{f^2}{\csf m}&0& -W_z
  \end{pmatrix}}_{:=M(t)}
  \begin{pmatrix}
      A_1\\
    A_2\\
     A_3
 \end{pmatrix}
  \end{aligned}
\end{equation}
where we used \eqref{relation} to simplify the expressions in the matrix $M(t)$, as well as \eqref{m} to write
\begin{equation}
    \frac{f^2m^2}{f^2m^2+k^4\csf^2}=\frac{f^2 k^4\csf^2}{k^2\csf^2-f^2}\frac{1}{\frac{f^2 k^4\csf^2}{k^2\csf^2-f^2}+k^4\csf^2}=\frac{f^2}{k^2\csf^2}.
\end{equation}
Multiplying the last equation of \eqref{50} by $\frac{fm}{k\csf^2}$ and summing it to the second one we get
\begin{equation}
    \frac{d A_2(t)}{dt}=-\frac{fm}{k^2\csf}  \frac{d A_3(t)}{dt},
\end{equation}
which gives
\begin{equation}\label{A2A3}
    A_2(t)=-\frac{fm}{k^2\csf}  A_3(t)
\end{equation}
if we choose the initial condition $A_2(0)=A_3(0)=0$, and, recalling the constraint \eqref{constraint}
\begin{equation}
    \A_0\cdot \boldsymbol{\xi}_0=0,
\end{equation}
with \begin{equation}
    \boldsymbol{\xi}(t)=\left(0,\, \frac{{k}}{m},\, \frac{f}{k\csf}\right)\qquad \forall\, t\geq 0
\end{equation}
it follows that $ \A_0$ should have the form $(a_1,0,0)$ with $a_1\in\mathbb{R}$. The most natural choice is  $ \A_0=(1,0,0)$.
As the linear system \eqref{50} is non-autonomous due to the $t$-dependence of $\Theta$, we transform it into an autonomous one by rotating the canonical basis by an angle 
\begin{equation}
    \beta=-\frac{k\csf t}{2}
\end{equation}
about the vector $\left(0,\, \frac{{k}}{m},\, \frac{f}{k\csf}\right)$. The rotation matrix (due to the Rodrigues formula \citep{rotation}) is given by
\begin{equation}\label{R}
    \mathcal{R}(t)=\begin{pmatrix}
        \cos\beta\ & -\frac{f}{k\csf} \sin\beta& \frac{k}{m}\sin\beta\\
       \frac{f}{k\csf}\sin\beta &1 -\frac{f^2}{k^2\csf^2}(1-\cos\beta) & \frac{f}{\csf m}(1-\cos\beta)\\
        -\frac{k}{m} \sin\beta & \frac{f}{\csf m}(1-\cos\beta) & 1-\frac{k^2}{m^2}(1-\cos\beta)
    \end{pmatrix}
\end{equation}
with inverse
\begin{equation}\label{RT}
    \mathcal{R}^{-1}(t)=\mathcal{R}^T(t)=\begin{pmatrix}
        \cos\beta\ & \frac{f}{k\csf} \sin\beta& -\frac{k}{m}\sin\beta\\
       -\frac{f}{k\csf}\sin\beta & 1 -\frac{f^2}{k^2\csf^2}(1-\cos\beta) & \frac{f}{\csf m}(1-\cos\beta)\\
        \frac{k}{m} \sin\beta & \frac{f}{\csf m}(1-\cos\beta) & 1-\frac{k^2}{m^2}(1-\cos\beta)
    \end{pmatrix}.
\end{equation}
The components $(Q_1(t), Q_2(t), Q_3(t))$ of the vector $\A(t)$ in the new basis are related to the components in the canonical basis, namely $(A_1(t), A_2(t), A_3(t))$, by
\begin{equation}\label{56}
    \begin{pmatrix}
        Q_1(t)\\
        Q_2(t)\\
        Q_3(t)
    \end{pmatrix}=\mathcal{R}^{-1}(t)    \begin{pmatrix}
        A_1(t)\\
        A_2(t)\\
        A_3(t)
    \end{pmatrix}.
\end{equation}
Even if  $(Q_1(t), Q_2(t), Q_3(t))$ and $(A_1(t), A_2(t), A_3(t))$ represent the same vector $\A(t)$ in two different basis, with abuse of notation we write $\Q(t)=(Q_1(t), Q_2(t), Q_3(t))$ and $\A(t)=(A_1(t), A_2(t), A_3(t))$. From \eqref{A2A3} and \eqref{56} it follows that
\begin{equation}\label{Q2Q3}
    Q_2(t)=-\frac{fm}{k^2\csf}  Q_3(t) \qquad\text{for all}\ t\geq 0.
\end{equation}
Differentiating with respect to time \eqref{56} and making use of \eqref{50} we get
\begin{equation}\label{system}
    \frac{d \Q}{dt}=\underbrace{\left[\frac{d \mathcal{R}^{T}}{dt}\mathcal{R}(t) + \mathcal{R}^{T}(t) M(t)\mathcal{R}(t) \right]}_{:= \mathcal{E}}\Q(t)
\end{equation}
where we used the fact that $\mathcal{R}^{-1}(t) =\mathcal{R}^{T}(t) $, and, due to \eqref{Q2Q3} we can rewrite \eqref{system} as the following planar system
\begin{equation}\label{sys red}
    \begin{pmatrix}
        \frac{dQ_1}{dt}\\
        \frac{dQ_3}{dt}
    \end{pmatrix}=\underbrace{\begin{pmatrix}
       \mathcal{E}_{11} & -\frac{fm}{k^2\csf} \mathcal{E}_{12}+\mathcal{E}_{13}\\
                \mathcal{E}_{31} & -\frac{fm}{k^2\csf} \mathcal{E}_{32}+\mathcal{E}_{33}
    \end{pmatrix}}_{:=\Pi} \begin{pmatrix}
        Q_1\\
        Q_3
    \end{pmatrix}.
\end{equation} Taking the time derivative of the matrix \eqref{RT} leads to
\begin{equation}
   \frac{d\mathcal{R}^T}{dt}=\begin{pmatrix}
       \frac{k\csf}{2} \sin\beta\ & -\frac{f}{2} \cos\beta& \frac{k^2\csf}{2m}\cos\beta\\
     \frac{f}{2} \cos\beta & \frac{f^2}{2k\csf}\sin\beta & -\frac{fk}{2m}\sin\beta\\
     -  \frac{k^2\csf}{2m}\cos\beta & -\frac{fk}{2m}\sin\beta & \frac{k^3\csf}{2 m^2}\sin\beta
    \end{pmatrix},
\end{equation}
and, consequently, 
\begin{equation}\label{RR}
   \frac{d \mathcal{R}^T}{dt} \mathcal{R}(t)=\begin{pmatrix}
      0\ & -\frac{f}{2} &  \frac{k^2\csf}{2m} \\
     \frac{f}{2}  &0 & 0\\
       - \frac{k^2\csf}{2m}  & 0 & 0
    \end{pmatrix}.
\end{equation}
Moreover, denoting by $S(t)=\mathcal{R}^{T}(t)M(t)\mathcal{R}(t)$, we have
\begin{equation}\label{E}
\begin{aligned}    S_{11}    &=\frac{km\csf a }{\Delta}e^{-ms}\sin(kq),\\
    -\frac{fm}{k^2\csf} S_{12}+S_{13}   &
    =\frac{m^2\csf a}{\Delta}e^{-ms}\cos(kq)-\frac{f^2 m}{k^2\csf}-\frac{m^3\csf a^2}{\Delta}e^{-2ms},\\
     S_{31}    & =\frac{k^2\csf a}{\Delta}e^{-ms}\cos(kq)+\frac{f^2 }{\csf m}+\frac{k^2\csf m a^2}{\Delta}e^{-2ms},\\
      -\frac{fm}{k^2\csf} S_{32}+S_{33}   &=-\frac{km\csf a }{\Delta}e^{-ms}\sin(kq),
\end{aligned}
\end{equation}
Summing up, due to \eqref{RR} and \eqref{E}, the planar system \eqref{sys red} reads as
\begin{equation}\label{reduced explicit}
\begin{pmatrix}
        \frac{dQ_1}{dt}\\
        \frac{dQ_3}{dt}
    \end{pmatrix}=
    \underbrace{\begin{pmatrix}
\dfrac{km\csf a}{\Delta}e^{-ms}\sin(kq) &
\begin{aligned}
\frac{m^2\csf a}{\Delta}e^{-ms}\cos(kq)
&+ \frac{k^2\csf^2 m - 2f^2 m}{2 k^2\csf} \\
&- \frac{m^3\csf a^2}{\Delta}e^{-2ms}
\end{aligned}
\\[6pt]

\begin{aligned}
\frac{k^2\csf a}{\Delta}e^{-ms}\cos(kq)
&+ \frac{2f^2 - k^2\csf^2}{2\csf m} \\
&+ \frac{k^2\csf m a^2}{\Delta}e^{-2ms}
\end{aligned}
&
-\dfrac{km\csf a}{\Delta}e^{-ms}\sin(kq)
\end{pmatrix}}_{:=\Pi}
\begin{pmatrix}
        Q_1\\
        Q_3
    \end{pmatrix}.
\end{equation}
Since $\mathrm{Tr}(\Pi)=0$,  
the eigenvalues of $\Pi$ are given by
\begin{equation}
    \lambda_{\pm}=\pm\sqrt{-\mathrm{Det}(\Pi)},
\end{equation}
with
\begin{equation}
    \mathrm{Det}(\Pi)=-\frac{k^2m^2\csf^2 a^2}{\Delta}e^{-2ms}+\left(\frac{k^2\csf^2-2f^2}{2k\csf}\right)^2,
\end{equation} leading to
\begin{equation}
    \lambda_{\pm}=\pm\sqrt{\frac{k^2m^2\csf^2 a^2}{\Delta}e^{-2ms}-\left(\frac{k^2\csf^2-2f^2}{2k\csf}\right)^2};
\end{equation}
consequently, as the rotation matrix $\mathcal{R}(t)$ is time-periodic, the time evolution of the amplitude vector $\A(t)$ is determined by the eigenvalues of the matrix $\Pi$. If $\lambda_+$ is real and positive, an exponential growth of $\A(t)$ will occur, with growth rate $\lambda_+$. Recalling that $\Delta=1-m^2a^2e^{-2ms}>0$, $\lambda_+$ is real and positive if and only if
\begin{equation}\label{>}
  m^2a^2e^{-2ms}>\frac{(k^2\csf^2-2f^2)^2}{ \left[4 k^4\csf^4 + \left(k^2\csf^2-2f^2\right)^2\right] }.
\end{equation}
The wave height in \eqref{Lag} and \eqref{LagA} is $2a\, e^{-ms}$, and the wave steepness, defined as the amplitude of the wave multiplied by the wave number $k$, is equal to $ka\, e^{-ms}$. Recalling the definition of $m$ in \eqref{m}, the inequality \eqref{>} reads as
\begin{equation}\label{relation2}
  k^2a^2e^{-2ms}> \frac{k^2\csf^2-f^2}{k^2\csf^2}\frac{(k^2\csf^2-2f^2)^2}{ \left[4 k^4\csf^4 + \left(k^2\csf^2-2f^2\right)^2\right] }.
\end{equation}
Therefore, if the square of the steepness of the waves described in \eqref{Lag} and \eqref{LagA} satisfies \eqref{relation2}, these are unstable, as a small perturbation of the flow will grow exponentially. One of the key features of the nonlinear waves solution \eqref{Lag} and \eqref{LagA} developed in \cite{mio} is its simple and explicit dispersion relation \eqref{dispersionC} (conversely to the $2$-layers model, where the dispersion relation is not explicit and complicated, see e.g. \citet{17,43,44}). Inserting the dispersion relation (c.f. \eqref{dispersionC})
\begin{equation}
\csf^2= \frac{f^2}{k^2}+\frac{f^4\csf_0^2}{\mathfrak{g}^2k^2}=\frac{f^2}{k^2}\left(1+\frac{f^2\csf_0^2}{\mathfrak{g}^2}\right)
\end{equation} 
into \eqref{relation2} we get the following condition for the instability
\begin{equation}\label{relation3prima}
  k^2a^2e^{-2ms}> \frac{f^2\csf_0^2}{(\mathfrak{g}^2+f^2\csf_0^2)}\frac{(f^2\csf_0^2-\mathfrak{g}^2)^2}{ \left[5f^4\csf_0^4+5\mathfrak{g}^4+ 6f^2\csf_0^2\mathfrak{g}^2\right] },
\end{equation}
which leads to
\begin{equation}\label{relation3}
  kae^{-ms}> \frac{f\left|\csf_0(f^2\csf_0^2-\mathfrak{g}^2)\right|}{\sqrt{(\mathfrak{g}^2+f^2\csf_0^2) \left[5f^4\csf_0^4+5\mathfrak{g}^4+ 6f^2\csf_0^2\mathfrak{g}^2\right] }}:=\mathscr{T}(\mathfrak{g},\csf_0),
\end{equation}
with the right-hand side (denoted by $\mathscr{T}$) depending only on physical constants  $f=2\Omega\approx 1.46\cdot10^{-4}\,\mathrm{s^{-1}}$ and $g=9.81\, \mathrm{ms^{-2}} $ and characteristics of the flow: $\csf_0$, and $\rho_0$, $\rho_1$ and $\rho_2$ (via $\mathfrak{g}=\left(\frac{\rho_1-\rho_0}{\rho_0}\right) \frac{\rho_2}{\rho_1}g$). See Fig. \ref{fig:both_figures} for plots of $\mathscr{T}$ as a function of $\mathfrak{g}$ and $\csf_0$.\\
In conclusion, a criterion for the instability of the proposed nonlinear $3-$layer model has been established, allowing for the determination of the occurrence of instability from the characteristics of the water column. In the following section, we provide some qualitative and quantitative insights and relations with observations in the Arctic Ocean.

\section{Discussion}\label{discussion}
We now conclude with some considerations about the instability threshold in \eqref{relation3} and the onset of instability. Let us start by analysing the behaviour of the  instability condition \eqref{relation3}
\begin{equation}
  kae^{-ms}> \mathscr{T}(\mathfrak{g},\csf_0),
\end{equation}
with respect to variations in $|\csf_0|$ and $\mathfrak{g}$. From Fig. \ref{fig:threshold} it is also possible to observe how the threshold $\mathscr{T}$ varies according to $|\csf_0|$ and $\mathfrak{g}$: an increase in $|\csf_0|$ gives an increase in $\mathscr{T}$, with such relation being connected with the value of $\mathfrak{g}$. For $\mathfrak{g}\approx0.005\,\rm{ms^{-2}}$, an increase of $|\csf_0|$ from approximately $0$ to $0.15\,\rm{ms^{-1}}$ leads the threshold $\mathscr{T}$ to increase from the approximate value of $10^{-5}$ to around $10^{-3}$, whereas the same increase in $|\csf_0|$ for $\mathfrak{g}\approx0.02\,\rm{ms^{-2}}$ imply an increasing of $\mathscr{T}$ from around $10^{-5}$ to around $10^{-4}$ (see Fig. \ref{fig:threshold} and Fig. \ref{fig:threshold2}). Conversely, increasing $\mathfrak{g}$ implies a decrease of $\mathscr{T}$.

Variations in $|\csf_0|$ and $\mathfrak{g}$ do also affect the wave steepness. In fact, by coupling \eqref{m} and \eqref{dispersionC}, we can write
\begin{equation}\label{m2}
    m^2=\frac{ \mathfrak{g}^2 k^4\csf^2}{f^4\csf_0^2}=\frac{ \mathfrak{g}^2 k^2}{f^2\csf_0^2}\left(1+\frac{f^2\csf_0^2}{\mathfrak{g}^2}\right)\approx\frac{ \mathfrak{g}^2 k^2}{f^2\csf_0^2},
\end{equation}
giving
\begin{equation}\label{m app}
    m\approx\frac{ \mathfrak{g}}{f|\csf_0|}k
\end{equation}
and consequently the following approximation for the wave steepness
\begin{equation}\label{approx steep
} kae^{-ms}\approx  kae^{-\frac{ \mathfrak{g}}{f|\csf_0|}k\,s}
\end{equation}
from which we observe that the wave steepness increases if $|\csf_0|$ increases, while it decreases if $\mathfrak{g}$ increases (holding all the other parameters fixed). Therefore, both the wave steepness and the instability threshold exhibit the same qualitative trend, increasing or decreasing depending on $|\csf_0|$ and $\mathfrak{g}$. Unfortunately, owing to their complicated functional dependence—arising from the large number of parameters involved in the wave steepness and from the complexity of the threshold expression—it is not possible to draw a simple, explicit conclusion. For instance, one cannot generally state that an increase in $\csf_0$ necessarily leads to a higher likelihood of instability, since this ultimately depends on the relative growth rates of $k a e^{-m s}$ and $ \mathscr{T}(\mathfrak{g},\csf_0)$ with respect to $\csf_0$.\\
It has been observed that, for internal waves in the ocean, the wave number $k$ is usually in the range between $10^{-1}$ cpm and $10^{-3}$ cpm (with the most energetic waves with $k$ between $10^{-2}$ cpm and $10^{-3}$ cpm, see \cite{20,DAsaroM}), corresponding to $2\pi\cdot10^{-1}\,\rm{m}^{-1}$  and $2\pi\cdot10^{-3}\,\rm{m}^{-1}$, respectively and that, in the Arctic Ocean, their amplitude is between $0.5\,\rm{m}$ and $2.5\,\rm{m}$ (see \cite{Cole_Toole_Rainville_Lee}) at depths between $70\,\rm{m}$ and $240\,\rm{m}$ \footnote{To be precise, the data in \cite{Cole_Toole_Rainville_Lee} are related to internal waves in a region of the Canada Basin, at around $75^\circ\,\rm{N}$. However, as we are interested in the order of magnitude of the amplitude, we elect to use these results for our analysis.}. These consideration allow us to infer that a plausible estimate for the wave steepness at the halocline base is
\begin{equation}\label{steep approx}
    \pi\cdot10^{-3}\lesssim kae^{-ms}\lesssim \frac{\pi}{2}.
\end{equation}
Moreover, the explicit dispersion relation \eqref{dispersionC} allows to easily compute the threshold as a function of the water properties: the densities of the three layers and the mean current above the halocline. With this scope, we analyse different salinity and temperature data in the central part of the Arctic Ocean, which can be found in \cite{54,58,SteeleETAL,A_C_C,W_C_A_S,57,42}, to provide some insights for different scenarios.\\
We denote the SML by water density $\rho_0$, potential temperature $\mathfrak{T}_0$, and salinity $\mathfrak{S}_0$; the halocline by water density $\rho_1$, potential temperature $\mathfrak{T}_1$, and salinity $\mathfrak{S}_1$; and the bottom layer of AW by water density $\rho_2$, potential temperature $\mathfrak{T}_2$, and salinity $\mathfrak{S}_2$. Density variations are then computed by (see \citep{57})
\begin{equation}\label{densityvariation}
    \frac{d\rho}{\rho}=-\alpha\, d\mathfrak{T}+\beta\, d\mathfrak{S},
\end{equation}
where  $\alpha\approx53\cdot10^{-6}\,\mathrm{K^{-1}}$ is the thermal expansion coefficient and $\beta\approx785\cdot10^{-6}\,\mathrm{kg}\,\mathrm{g^{-1}}$ is the haline contraction coefficient. Both $\alpha$ and $\beta$ depend on the water properties, therefore the values reported here refer only for the Arctic Ocean (see \cite{57}).\\
With the above formula \eqref{densityvariation}, it is possible to estimate the reduced gravity in \eqref{redg} as
\begin{equation}\begin{aligned}
     \mathfrak{g}& =g\left(\frac{\rho_1-\rho_0}{\rho_0}\right) \frac{\rho_2}{\rho_1}\approx g\left[-\alpha(\mathfrak{T}_1-\mathfrak{T}_0)+\beta(\mathfrak{S}_1-\mathfrak{S}_0)\right]\left\{1+\left[-\alpha(\mathfrak{T}_2-\mathfrak{T}_1)+\beta(\mathfrak{S}_2-\mathfrak{S}_1)\right]\right\},
\end{aligned}   
\end{equation}
and, recalling that $f=2\Omega\approx 1.46\cdot10^{-4}\,\mathrm{s^{-1}}$, it is therefore possible to compute the value of the instability threshold  $\mathscr{T}( \mathfrak{g},\csf_0)$ for different configurations of the Arctic Ocean, according to the formula \eqref{relation3}. In Table \ref{tab:watervalues} the value of $\mathscr{T}$ is reported, for data in \cite{54,57,58,SteeleETAL,CoachmanAagaard,A_C_C,W_C_A_S,42}\footnote{\label{fn:shared}Actually, in \cite{A_C_C,CoachmanAagaard} the data report the temperature of the water and not the potential temperature. As our intent is to provide some quantitative results of the instability, and, as it is shown in \cite{57}, potential temperature and temperature basically are the same for water above $1000\,\rm{m}$ of depth, we use this temperature data as if the are potential temperature} for $|\csf_0|=0.1\,\rm{ms^{-1}}$. 
\begin{table}
    \centering
\begin{tabular}{c|c|c|c|c|c|c|c|c|c|}
{} &
\rotatebox{90}{\cite{54}} &
\rotatebox{90}{\cite{57}} &
\rotatebox{90}{\cite{SteeleETAL}} &
\rotatebox{90}{\cite{58}} &
\rotatebox{90}{\cite{CoachmanAagaard}} &      
\rotatebox{90}{\cite{A_C_C}} & 
\rotatebox{90}{\cite{W_C_A_S}} &
\rotatebox{90}{\parbox{4cm}{ \cite{42}\\ January 2014}} &
\rotatebox{90}{\parbox{4cm}{ \cite{42}\\ July 2014}} \\
\hline
        $\mathfrak{T}_0$ [$^\circ$ C]& $-1.5$  & $-1.7$& $-1.5$& $-1.5$& $-1.7$& $-2$ & $-1.5$  & $-2$ & $-2$\\
        $\mathfrak{S}_0$  [psu] & $34.0$ & $32$ & $29$ & $31$ &$34$ & $33.5$ & $31$ & $32.5$ & $34$\\
         $\mathfrak{T}_1$ [$^\circ$ C] & $0$& $0.6$ & $-1.1$ & $-1.0$ & $0$ & $-2$ & $-1$ & $-0.5$ & $0$\\
        $\mathfrak{S}_1$ [psu]  & $34.2$  & $34.85$ & $32$ & $33$ & $34.4$ & $34.5$ & $33$& $34$ & $34.2$\\
         $\mathfrak{T}_2$ [$^\circ$ C] & $2$ &  $-0.5$ & $-0.8$ & $-0.5$ &$1$ & $0.7$ & $-0.5$ & $1.5$ & $2$\\
        $\mathfrak{S}_2$  [psu]  & $34.9$ & $34.9$ & $34$ &$35$& $34.5$ & $34.5$ & $34.5$ & $34.5$ & $34.5$ \\ \hline
         $\mathfrak{g}$ [$\mathrm{ms^{-2}}$] &$7.6\cdot10^{-4}$ &$2.1\cdot10^{-2}$ &$2.3\cdot10^{-2}$ &$1.5\cdot10^{-2}$ &$2.2\cdot10^{-3}$  &$7.7\cdot10^{-3}$ &$1.5\cdot10^{-2}$ &$1.1\cdot10^{-2}$ &$5\cdot10^{-4}$ \\ \hline
         $\mathscr{T}(\mathfrak{g}, |\csf_0|=0.1 \,\mathrm{\frac{m}{s}})$ & $8.6\cdot10^{-3}$ & $3.1\cdot10^{-3}$ & $2.9\cdot10^{-4}$ & $4.3\cdot10^{-4}$ & $3.0\cdot10^{-3}$ & $8.5\cdot10^{-4}$& $4.4\cdot10^{-4}$ & $5.9\cdot10^{-4}$ & $1.3\cdot10^{-2}$
    \end{tabular}
    \caption{Water properties from various sources in the literature, along with the corresponding computed values of $\mathfrak{g}$ and $\mathscr{T}$, for $|\csf_0|=0.1 \,\mathrm{ms^{-1}}$.}
    \label{tab:watervalues}
\end{table}\\
Comparing the estimates of the wave steepness \eqref{steep approx} with the estimates for the threshold $\mathscr{T}$ in Table \ref{tab:watervalues} suggests that instability is likely to occur at the bottom of the halocline. As instability is widely recognized as a mechanism contributing to mixing processes (see, e.g., \cite{Staquet_CISM, S_S, KH_instab}), the proposed model may offer insight into the processes associated with the degradation of the halocline reported in \cite{51, Morison2018}. Specifically, instability occurring near the base of the halocline may facilitate mixing between halocline waters and the underlying warmer and saltier Atlantic Water.\\
Before proceeding further, let us also give an estimate of the magnitude of $m$, from \eqref{m app}. To this end, fix $|\csf_0|\approx 0.1\, \mathrm{ms^{-1}}$, and recall that $f=2\Omega\approx 1.46\cdot10^{-4}\,\mathrm{s^{-1}}$. We then choose $k$ between $10^{-2}$ cpm and $10^{-3}$ cpm, and, referring to Tab. \ref{tab:watervalues}, the values of $\mathfrak{g}$ are between $7.5\cdot10^{-4}\, \rm{ms^{-2}}$ and $2.3\cdot10^{-2}\, \rm{ms^{-2}}$. Consequently, 
\begin{equation}
    0.3\lesssim m\lesssim 100\quad [\mathrm{m^{-1}}].
\end{equation}
As above we have established that instability is likely to occur at the halocline base, and as the model presented here describes waves whose amplitude, hence steepness, increases with depth, recalling that instability is related to the wave steepness, a natural question that arise is what happen to shallower waves. Recalling that we denoted by $\mathrm{s}_-$ the halocline base, and denoting by $\delta$ the height above the halocline base of the wave under consideration, we see that the ratio between the steepnesses of the shallower wave and of the one describing the halocline base is 
\begin{equation}
    \frac{ka\,e^{-m(\mathrm{s}_-+\delta)}}{ka\,e^{-m\mathrm{s}_-}}=e^{-m \delta}.
\end{equation}
For $\delta = 10\,\mathrm{m}$, we have $e^{-m\delta} < 0.05$; therefore, even though the instability condition \eqref{relation3} is only a sufficient condition for the onset of instability (that is, other more general conditions may also lead to instability, and a wave whose steepness does not satisfy \eqref{relation3} is not necessarily stable), we may infer that, within the proposed model, the top of the halocline is unlikely to be unstable.\\
In conclusion, in the present work we provided an instability analysis of the nonlinear near-inertial Pollard wave model for the central Arctic Ocean (modeled as a 3-layer ocean) proposed in \cite{mio}. Two of its features are particularly important: firstly, this solution recover, even with some simplification, the trend of the halocline's upper surface to become shallower when moving from the Amerasian basin to the Eurasian Basin, and secondly, due to its simple dispersion relation, allow us to compute an instability condition based on the wave steepness and property of the flow (densities and mean current). Based on the wave-amplitude measurement of \cite{Cole_Toole_Rainville_Lee} and from the water properties in \cite{54,57,58,SteeleETAL,CoachmanAagaard,A_C_C,W_C_A_S,42} we have been able to infer that the bottom of the halocline is likely to be unstable, a condition that can lead to the mixing---in the lower part of the halocline---of the colder and fresher halocline water with the saltier and warmer Atlantic Water, contributing to the halocline weakening observed in \cite{51}.

\acknowledgments
The author is supported by the Austrian Science Fund (FWF) [grant number Z 387-N, grant doi \url{https://doi.org/10.55776/Z387}].\\
The author declares no conflicts of interest.

\datastatement
No data were created for this paper. All data used in this work are properly cited and referred to:   \cite{54,57,SteeleETAL,58,CoachmanAagaard,A_C_C,W_C_A_S,42}.



\bibliographystyle{ametsocV6}

\bibliography{bib2}

\end{document}